\documentclass[osajnl,showpacs,10pt]{revtex4}
\usepackage{amsmath,amssymb,graphicx}
\usepackage{float}
\usepackage{multirow}
\usepackage{colortbl}
\DeclareMathOperator{\sech}{sech}
\usepackage[section]{placeins}
\begin{document}
	\title{Adding Binary Numbers with Discrete Solitons in Waveguide Arrays}
	\author{Alaa Shaheen, Amaria Javed, and U.~Al Khawaja\\
		Department of Physics, United Arab Emirates University,\\ P.O. Box
		15551, Al-Ain, United Arab Emirates.\\ u.alkhawaja@uaeu.ac.ae}
	\date{}
\begin{abstract}
 We present a design and protocol to add binary numbers using discrete solitons in waveguide arrays. We show that the nonlinear interaction between discrete solitons in waveguide arrays can be exploited to design half and full adders. By modulating the separation between waveguides and introducing control solitons, we achieve the performance of an XOR gate. We construct the half and full adders using the XOR gate together with the previously- designed OR and AND gates. To facilitate the experimental realization, we calculate the profile of separations between the waveguides that will lead to the performance of the XOR gate.
\end{abstract}
	\maketitle
\section{Introduction} \label{introsec}
The increasing demand on higher speed and capacity of data processing motivates replacing current electronic data processing by optical data processing in an analogy with the successful replacement of electronic data communication by optical data communication \cite{books1, books2, books3, books4}.  In a quest to achieve a comprehensive optical data processing, many of the main ingredients have been already proposed in the literature considering various setups \cite{gates1,gates2,gates3,gates4,gates5,gates6,gates7,gates8,gates9,gates10,gates11,gates12,gates13}. Prominent among these is discrete solitons in waveguide arrays \cite{gates1,gates2}.  Optical solitons have been suggested as data carriers due to their unique feature of preserving their integrity over long distances of propagation and their particle-like inelastic scattering with each other and with external potentials \cite{books5,fedor1,fedor2,fedor3,usamanjp}. Discrete solitons are also characterised by this appealing feature. In addition, the propagation of discrete solitons in waveguide arrays can be most feasibly controlled through the dispersion or nonlinearity management \cite{andrey,lepri}.
Dispersion management is achieved by varying the separations between the waveguides resulting in an effective potential \cite{usamaandrey}.  The profile of separations' variations can be set such that a particular type of reflectionless potential is realized \cite{andrey}. In such a case, the soliton will scatter off the potential elastically with minimized radiation losses. This results in clean particle-like soliton scattering. The nonlinear interaction together with the dispersion management allow for a host of setups where various data processing components can be devised. For instance, unidirectional flow has been shown to exist in waveguide arrays with dispersion management \cite{usamaandrey} and nonlinearity management \cite{submitted} in a similar manner as for the continuum cases in optical fibers with double potential wells \cite{usamaasad}, PT-symmetric potentials \cite{usamayuri}, and nonlinearity management \cite{recentpre}.\\
\\Using the same dispersion management method, logic gates have been proposed in Ref. \cite{1}. Control solitons were introduced to propagate through the potential wells and were shown to be an effective tool to modulate the depth of the potential well. The output was then controlled by the intensity of the control solitons and the AND, OR, NAND, and NOR logic gates were all achieved with a single devise but different protocols. Here, we follow the same procedure building on this previous work and achieve the XOR gate which is instrumental in realising half and full adders of binary number. The XOR gate is designed out of a modification on the OR gate where we introduce a third potential well and control soliton that further disperses the outputs of the OR gate based on small center-of-mass speed difference. When a soliton scatters off the potential wells it will suffer from a reduction in its center-of-mass speed due to exciting internal modes. The third potential well and control soliton are capable of detecting this reduction in speed and directing the soliton to an opposite direction compared with the case in OR gate. \\
\\The propagation of solitons in the proposed device is modeled by the discrete nonlinear Schr\"odinger equation in the tight-binding approximation with site-dependent dispersion coefficients. The potentials are introduced through pre-calculated profiles for the waveguides' separations. At first, the stationary solutions are obtained which were then used as control and signal solitons for the time evolution of solitons in the waveguide. Transport coefficients in terms of the soliton's speed are then calculated and the window in the velocity domain for which the device functions in the desired manner are identified. \\
\\	To construct half and full adders, a number of gates have to be connected, namely the output of a certain gate needs to be the input of another gate. Since the typical output of a gate is a reflected or transmitted signal soliton and the input is a control soliton, which are in general different in intensity, we invoke the concept of a converter. This is a pulse source that generates an output of standard intensity. It is triggered by an input pulse of any intensity. The intensity of the output pulse is a characteristic of the converter and is independent of that of the input pulse.  We believe such a pulse source is realisable. One-bit half adder can be connected in this manner to result in a one-bit full adder and, hence, one-bit full adder can be connected to obtain an $n$-bit full adder. In the present paper, we show only how one-bit half and one-bit full adders can be constructed since it is well-established how to obtain the $n$-bit adders from the one-bit adders.\\
\\The rest of the paper is organized as follows. In Section \ref{modsec}, we present the setup and theoretical model. In Section \ref{xorsec}, we show how the OR gate can be modified to generate an XOR gate. This section starts with a review of the setup and performance of the OR gate followed by the setup modification and performance of the XOR gate. In Section \ref{adderssec}, we show how the gates can be connected to give half and full adders. Finally, in Section \ref{concsec}, we summarize our main conclusions.
\section{Theoretical Framework and Setup} \label{modsec}
The propagation of solitons in a one-dimensional array of $N$ waveguides with focusing nonlinearity can be described, in the tight-binding approximation, by the following discrete nonlinear Schr\"odinger equation (DNLSE) \cite{rev},
	\begin{equation}\label{DNLSE}
		i\dfrac{\partial \Psi_n }{\partial z} + C_{n,n-1} \Psi_{n-1} + C_{n,n+1} \Psi_{n+1} +
		\gamma |\Psi_n|^2\Psi_n =0,
	\end{equation}
	where $\Psi_n$ is the normalized mode amplitude and $n$ is an 
	integer associated with the waveguide channel, $z$ is the
	propagation distance, $C_{n,m}$ are the coupling coefficients between different
	waveguide channels $n$ and $m$, and $\gamma$ is the strength of the focusing nonlinearity. Here we used $\gamma = 1$ for numerical simulation.\\
	\\The modulation of the coupling constants via changing the separation between waveguides
	leads to an effective potential and Eq.~\eqref{DNLSE} will be rewritten as \cite{andrey}:
	\begin{equation}
		i\dfrac{\partial \Psi_n}{\partial z} = - C_{n-1}^S \Psi_{n-1} - C_{n+1}^S \Psi_{n+1} -
		\gamma |\Psi_n|^2\Psi_n,
	\end{equation}
	where
	\begin{equation}\label{CM}
		C_{n\pm1}^S = \sqrt{(C+|{\Psi_{n}^{AL}|^2})(C+|{\Psi_{n\pm1}^{AL}|^2)}}.
	\end{equation}
	For the effective potential to be reflectionless,
	we use the integrable Ablowitz-Ladik model
	\cite{AL}
	\begin{equation}
		i\dfrac{\partial \Psi_n}{\partial z} + (\Psi_{n-1} + \Psi_{n+1})(C+|\Psi_n|^2)=0,
	\end{equation}
	with the exact soliton solution
	\begin{equation}
		\Psi_{n}^{AL}=\sqrt{c}\sinh(\mu)\sech[\mu(n-n_0)]\exp(i\beta z)\label{alsol},
	\end{equation}
	where $\beta = 2c \cosh(\mu)$, $\mu$ is the inverse width of the soliton, 
	$n_0$ corresponds to the location of the soliton peak, and $c$ is an arbitrary real constant. We refer to reflectionless as a clean scattering where all of the soliton intensity is either transmitted or reflected.\\
	\\The coupling strength between pulses in neighboring waveguides decays exponentially
	in terms of the waveguides separation, as \cite{2,expt2}
	\begin{equation}\label{Ceq}
		C_{n,n\pm1}^S=C\exp\Big(1-\dfrac{D_{n,n\pm1}}{D_0}\Big),
	\end{equation}
	where $D_{n,n\pm1}$ is the separation between waveguides $n$ and $n\pm1$, and
	$D_0$ and $C$ are positive constants. Thus, the separation between waveguides
	that gives rise to an effective potential is obtained by inverting the last equation,
	namely
	\begin{equation}\label{Deq}
		D_{n,n\pm1}=D_0\Bigg[1-\log\Bigg(\dfrac{C_{n,n\pm1}^S}{C}\Bigg)\Bigg].
	\end{equation}
	This practical relation is used to design specific effective potentials.\\
	\\All solitons used here are the stationary states of Eq.~\eqref{DNLSE}. They are generated using the Newton-Raphson method and trial solution is given by $\Psi_{n} = A \exp(-\alpha |n-n_0|)$, where $\alpha^{-1}$ and $n_0$ are parameters that set the width and peak location of the soliton respectively. As usual, two stationary modes result out of this procedure, namely, the Page mode and the Sievers-Takeno mode \cite{Boris}. Therefore, the initial soliton used in our protocol can be written generally as  
	\begin{equation}\label{initial}
	\Psi = \Psi_s \,e^{i v n} + r~\Psi_{c_1} + s~\Psi_{c_2}+ p~\Psi_{p},
	\end{equation}
	where $ \Psi_s$, $\Psi_{c_1}$, $\Psi_{c_2}$, and $ \Psi_p$ are signal soliton, input-1 control soliton, input-2 control soliton, and partitioner control soliton, respectively, which are generated by the scheme described above. Their profiles are given by the AL solution, Eq. (\ref{alsol}). The coefficient $e^{i v n}$ corresponds to the kick-in velocity of the signal soliton with velocity, $v$. The parameters $r$, $s$, and $p$ control the intensities of the control solitons. Typically we use $r=s=0.16$ and $p=0.1$.
	
The intensity of the control soliton is $\sech^2(x)$-shaped and the modulation of the dispersion coefficients is also $\sech^2(x)$-shaped. The control soliton is located at the region where the dispersion is modulated. Therefore, the incident soliton will encounter two effective potentials when it is scattered. The first comes from the modulation in the dispersion coefficients and the second comes from the nonlinear interaction with the control soliton. Since the two effective potentials have the $\sech^2(x)$ shape, the net interaction energy will be determined by the relative values of the strengths of the two potentials. By controlling the intensity of the control soliton, we tune the strength of the effective potential resulting from the modulation in dispersion coefficients.\\
	\\The consideration of input and output of the logic gate with this scheme is described in next section. The use of discrete solitons for the addition of binary numbers requires an XOR	gate which is designed in the next section by modifying our previous scheme for the OR gate \cite{1}.
	
\section{The XOR gate}
\label{xorsec}
The XOR gate we are proposing here is a variation of the OR gate by \cite{1}. We start in the next subsection by reviewing the OR gate and showing how it can be modified to function as an XOR gate.
\subsection{The OR gate}
\begin{figure}[!h]
		\centering
		\includegraphics[width=0.75\linewidth]{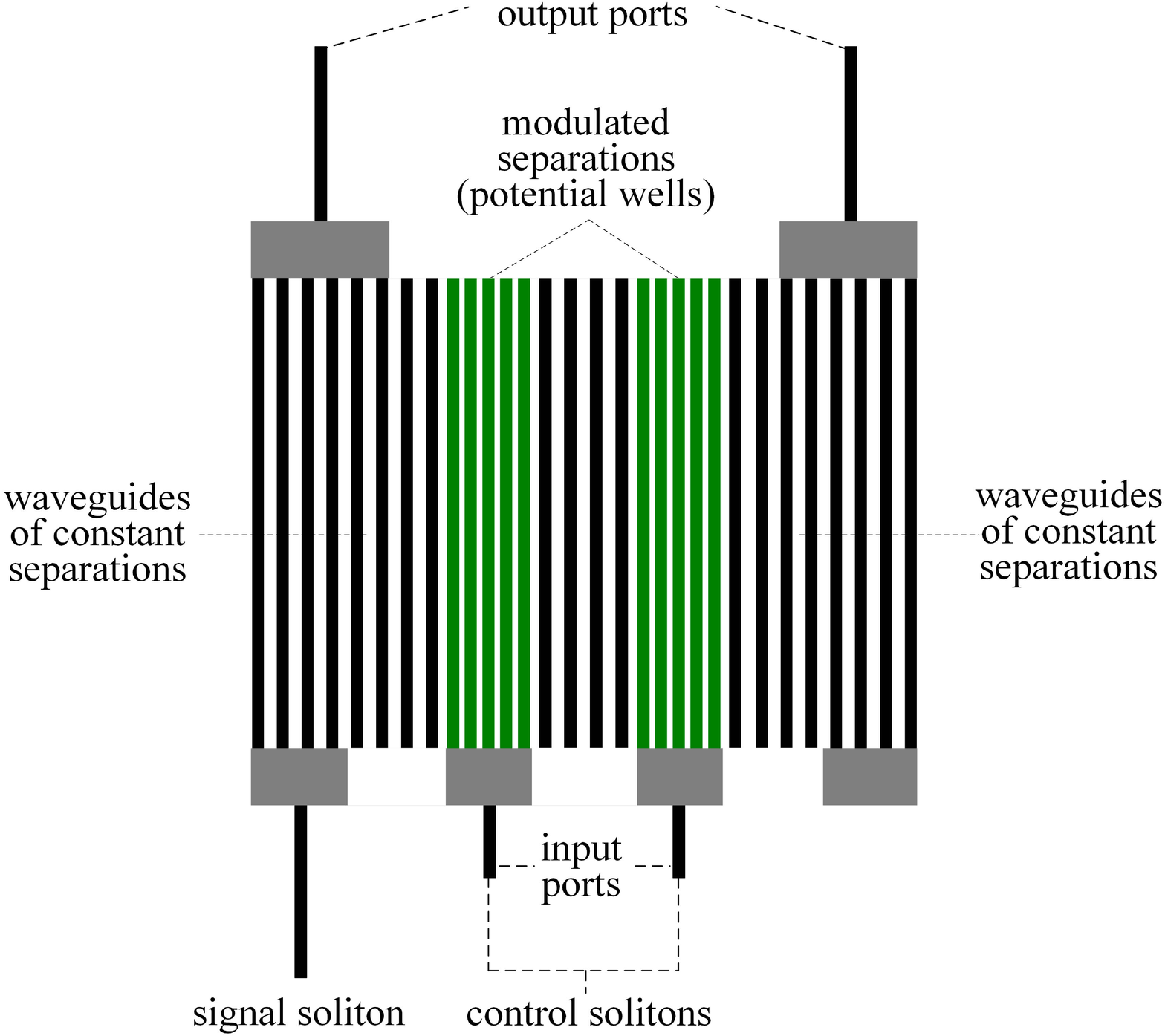}
		\caption{Schematic figure representing the OR gate having two potential wells injected with two control solitons.}
		\label{fig:schematic-OR}
	\end{figure}

We follow a similar scheme of controlling the scattering of the signal soliton by injecting a control soliton into a potential well as described in Ref. \cite{1}. The schematic figure representing the OR gate that includes two potential wells, is shown in Fig.~\ref{fig:schematic-OR}.
	An effective single potential well can be obtained from Eq.~\eqref{CM} with the modulated coupling constants
	\begin{eqnarray}
		C_{n\pm1}^S&=&C\{[1+\sinh^2(\mu)\sech^2\mu(n-n_0)]\nonumber\\
		&\times&[1+\sinh^2(\mu)\sech^2\mu(n\pm1-n_0)]\}^{1/2}
		\label{eq8}.
	\end{eqnarray}
	
\begin{figure}[!h]
\hspace*{-0.5cm} 
\centering
\includegraphics[scale=0.85]{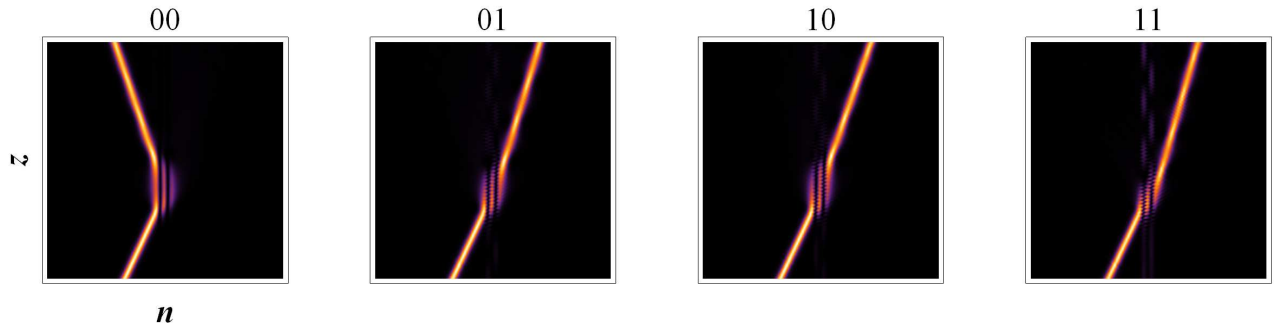}
\caption{Spacio-temporal plots representing the function of OR gate with initial soliton speed $v=0.22$. The two potential wells are separated by 4 waveguides, $\Delta n =4$, the amplitude of the control soliton is multiplied by the power control parameter $r=0.16$, as defined by Eq.~\eqref{initial} the  waveguide arrays range from $n=1$ to $n=124$ and time ranges from $z=0$ to $z=317$. Other parameters used are $c=1$ and $\mu=1.5$. For clarity purpose, the control solitons are not shown here.}
\label{fig:dp-OR}
\end{figure}

\begin{figure}[!h]
		\centering
		\includegraphics[width=0.8\linewidth]{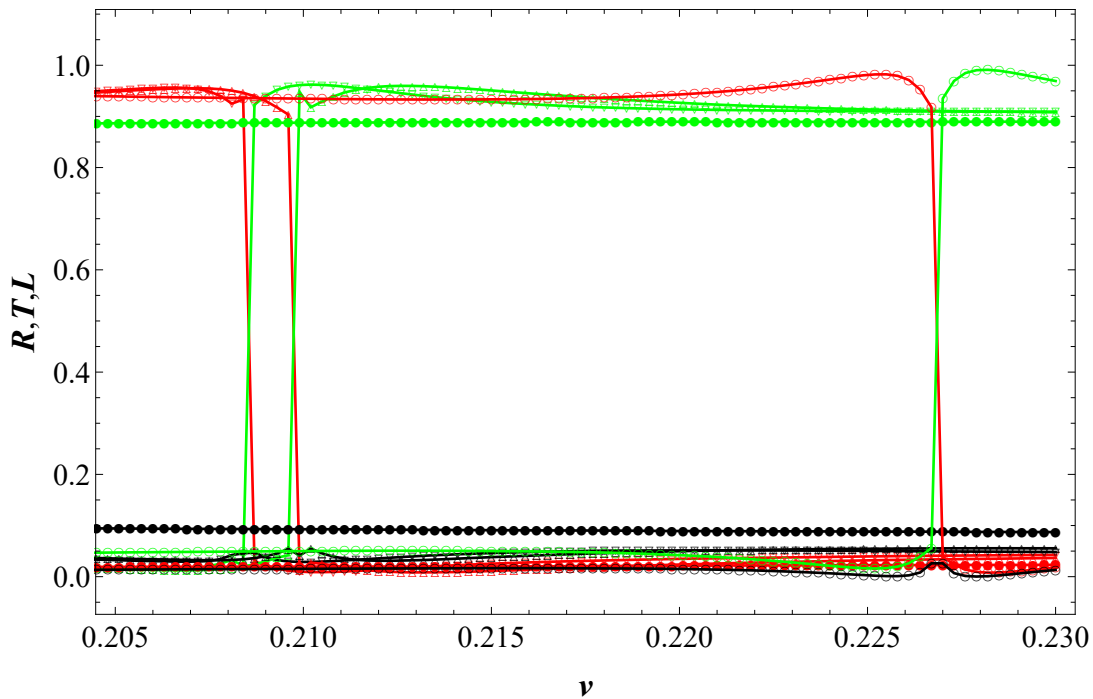}
		\caption{Transport Coefficients of OR gate. Red curves correspond to reflection (R), green curves correspond to transmission (T), and black curves correspond to trapping (L). Filled circles correspond to the presence of control solitons in both wells (11). Up and lower triangles correspond to the presence of a control soliton in the left or right well (10 or 01), respectively. Empty circles correspond to the absence of control solitons from both wells (00). All parameters used are the same as in Fig.~\ref{fig:dp-OR}.}
		\label{fig:TC-OR}
	\end{figure}

	This is achieved by the waveguides separations profile given by Eq.~\eqref{Deq}, namely
	\begin{eqnarray}
		D_{n,n\pm1}&=&D_0\Big[1-\dfrac{1}{2}\log[1+\sinh^2(\mu)\sech^2(\mu(n-n_0))]\nonumber\\
		&-&\dfrac{1}{2}\log[1+\sinh^2(\mu)\sech^2(\mu(n\pm1-n_0))]\Big].
	\end{eqnarray}
	Similarly, a double potential well is obtained by generalizing Eq. (\ref{eq8}) as follows
	\begin{eqnarray}
		C_{n,n\pm1}^S&=&[(C+|\Psi_{1,n}^{AL}|^2+|\Psi_{2,n}^{AL}|^2)\nonumber\\
		&\times&(C+|\Psi_{1,n\pm1}^{AL}|^2+|\Psi_{2,n\pm1}^{AL}|^2)]^{1/2},
	\end{eqnarray}
	where $\Psi_{1,2}^{AL}$ are two exact solitonic solutions centered at different waveguides which take the form
	\begin{eqnarray}
		\Psi_{i,n}^{AL}=\sqrt{c}\sinh(\mu_i)\sech[\mu_i(n-n_i)]\exp(i\beta_i z),&i=1,2,&
	\end{eqnarray}
	with $\beta_i = 2c \cosh(\mu_i)$, $\mu_i$ is the inverse width of the $i$-th soliton and $n_i$ corresponds to the location of the $i$-th soliton's peak.\\
To achieve the performance of OR gate, we solve Eq.~\eqref{DNLSE} using the above mentioned coupling coefficients. All solitons used at the start of the time evolution, whether signal or control solitons, are the stationary solitons of Eq.\eqref{DNLSE}. The input signals of the logic gate are taken as 1 if there is a control soliton in the potential well and 0 otherwise, such that 11 corresponds to two equal control solitons in the two potential wells, 01(10) corresponds to a control soliton in the right (left) potential well, and 00 corresponds to no control solitons in the potential wells. The output is taken from the scattered soliton which can be reflected, transmitted, or trapped. It is clear from the spacio-temporal plots shown in Fig.~\ref{fig:dp-OR} that for the 00 case the scattered soliton is reflected while for all other options of the input, the scattered soliton is transmitted. The performance of the OR gate is thus obtained if the output is taken from the transmission port. Hence, obeying the Boolean expression for the OR gate.\\
\\Now, we present the calculation of the transport coefficients which are defined as follows:\\
reflection $R=\sum_{1}^{n_1-\delta n}|\Psi_n|^2 /  \sum_{1}^{N}|\Psi_n|^2$, transmission $T = \sum_{n_2+\delta n}^{N}|\Psi_n|^2 /  \sum_{1}^{N}|\Psi_n|^2$ and trapping $ L = \sum_{n_1-\delta n}^{n_2+\delta n}|\Psi_n|^2 /  \sum_{1}^{N}|\Psi_n|^2 $, where $N$ is the number of waveguides and $\delta n$ is roughly equal to the width of the soliton in order to avoid the inclusion of the tails of the trapped soliton with the reflected or transmitted ones.\\
\\It should be noted that we use throughout all numerical calculated soltions that are in-phase with each other. We realize that phase differences may affect in the output of the scattering of solitons. Therefore, we chose for simplicity not to incorporate this additional factor in our protocol.
A preliminary investigation of the scattering outcomes in terms of the potential and soliton parameters including potential depth, width, location, soliton initial speed, phase, and type, gives an idea of the ranges of parameters for which the useful applications could be obtained. The transport coefficients curves for the OR gate are presented in Fig.~\ref{fig:TC-OR} which show a reasonably wide window of velocity for the operation of OR gate.
	 \subsection{The XOR Gate: A modified OR gate}
\begin{figure}[!h]
        \centering
            \includegraphics[scale=0.8]{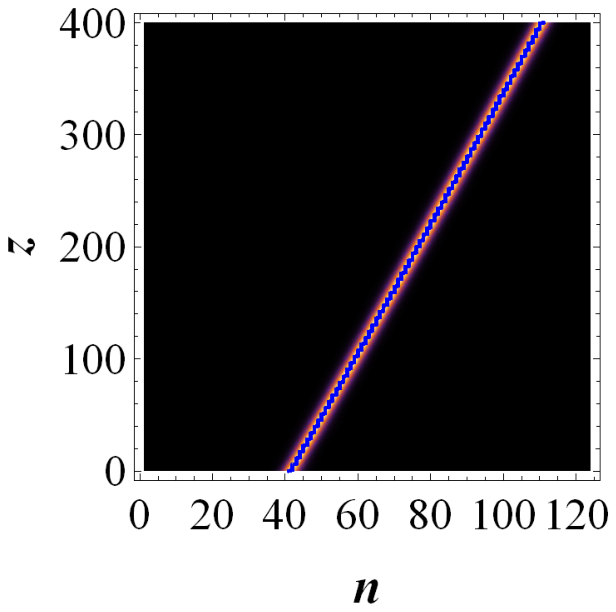}
            \includegraphics[scale=0.8]{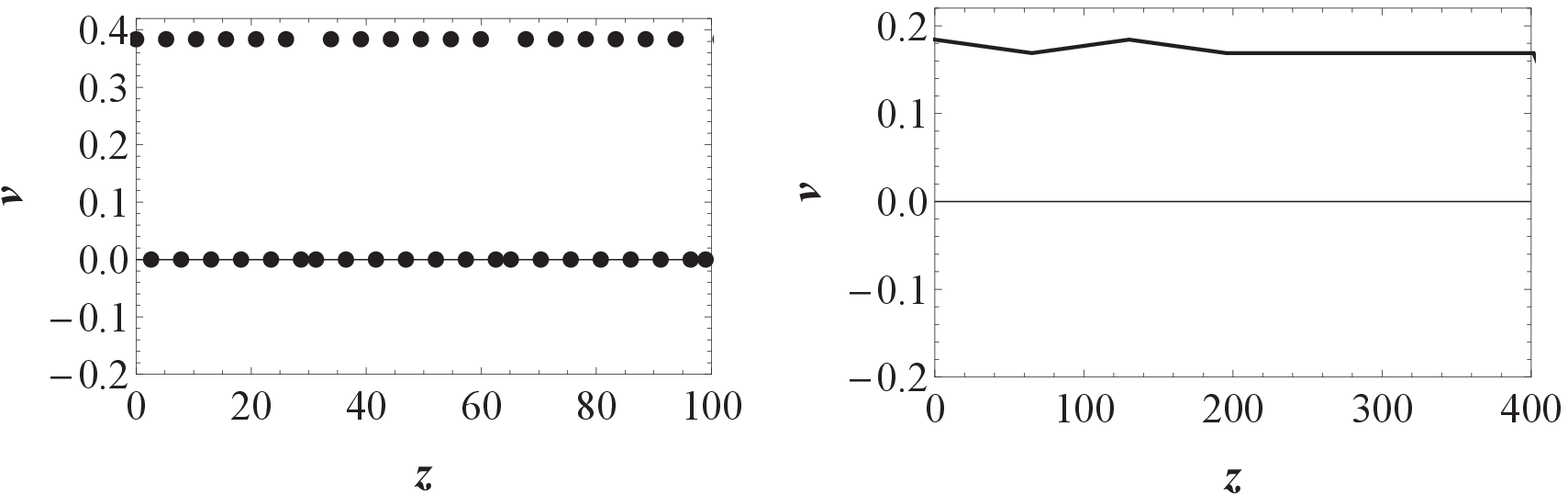}
        \caption{Spacio-temporal plot representing
            the free propagation of a signal soliton in the setup of OR gate with
            both potential wells and control solitons switched off.
            The trajectory of the signal soliton with initial  speed $v=0.214$
            is represented by blue line. The left subfigure in lower panel
            shows two possible values of speed, namely $v=0$ corresponding to soliton pinning,
            and $v\approx0.39$ corresponding to transition from site to site.
            The right subfigure in lower panel shows the averaging of the two
            speeds over several waveguides.
            Waveguides range from $n=1$ to $n=124$ and time ranges from $z=0$ to $z=400$. Other parameters used are $c=1$ and $\mu=1.5$.}
        \label{fig:benchmark}
    \end{figure}

\begin{figure}[!h]
        \centering
        \includegraphics[width=0.8\linewidth]{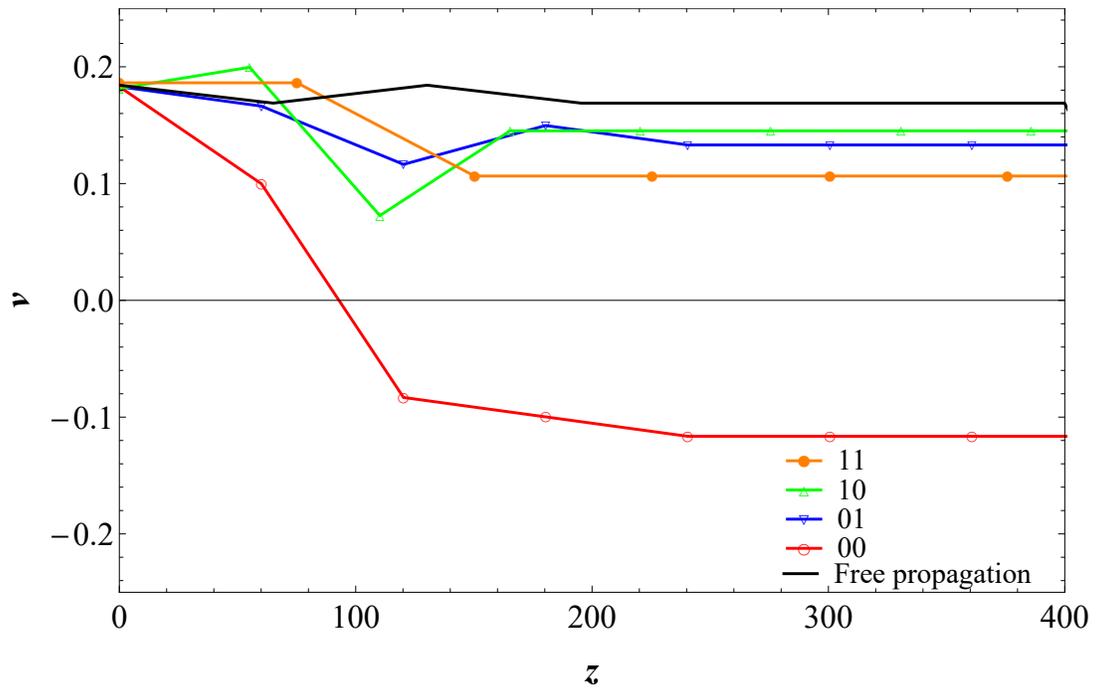}
        \caption{The effect of the potential wells on the speed of signal soliton in the OR gate.  All cases have same initial the speed $v=0.214$. The free propagation of signal soliton leads the highest state among all. The (00) corresponds to reflection. The speeds of the 01 and 10 are closer to each other and are noticeably larger than the (11) state. }
        \label{fig:speeds}
    \end{figure}

\begin{figure}[!h]
        \centering
        \includegraphics[width=0.75\linewidth]{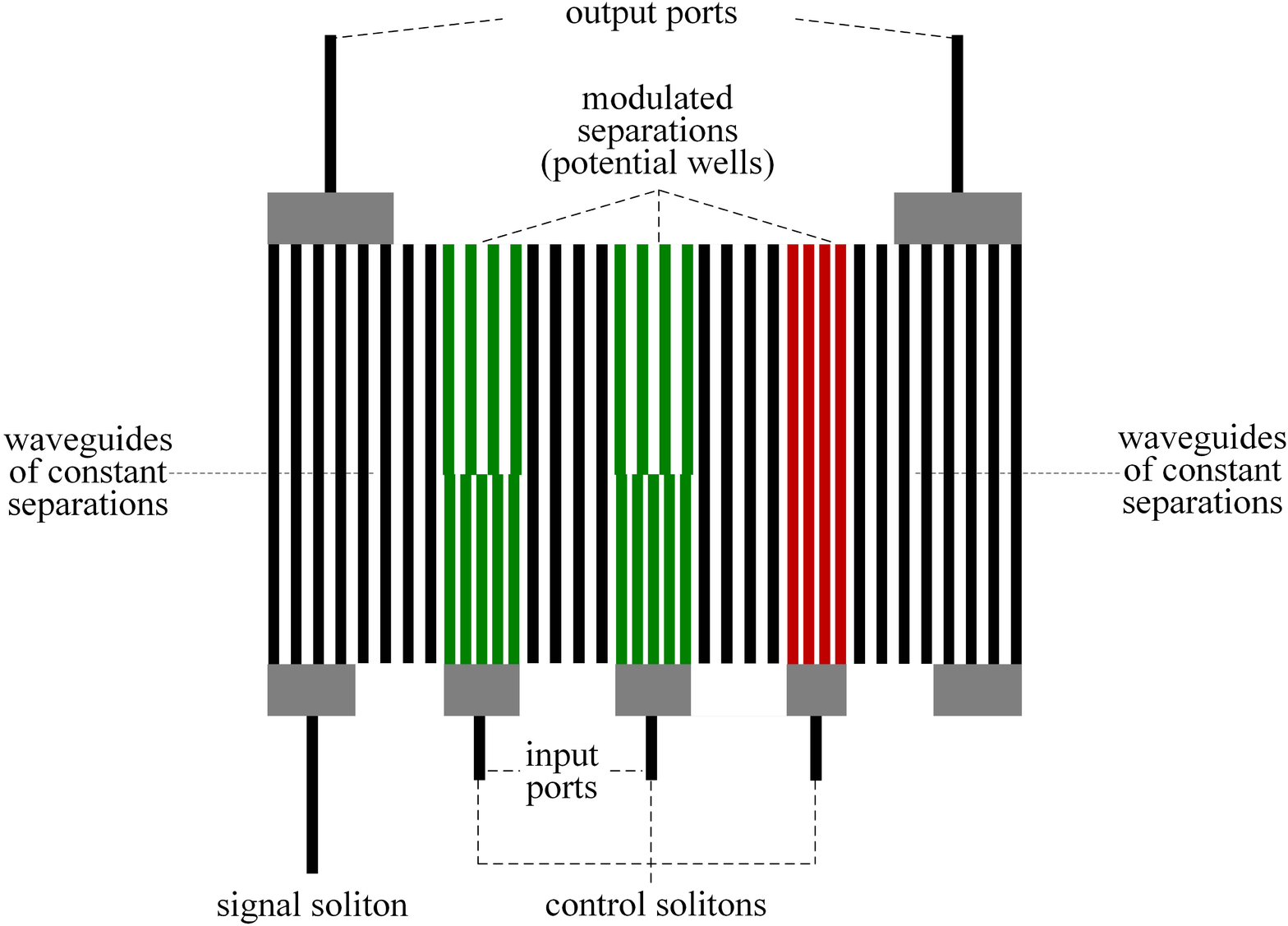}
        \caption{Schematic figure showing the XOR gate having two modified potential wells of the same control power and a third potential well is introduced with different control power
            to reflect the (11) case in order to achieve the performance of XOR gate.
            The first two potential wells extend only up to about
            half of the evolution time in order not to affect the reflected signal in the (11) case. Beyond this point, the separations return back to their uniform values away from the potentials.}
        \label{fig:schematic-XOR}
    \end{figure}

\begin{figure}[!h]
\hspace*{-0.8cm} 
\centering
\includegraphics[scale=0.85]{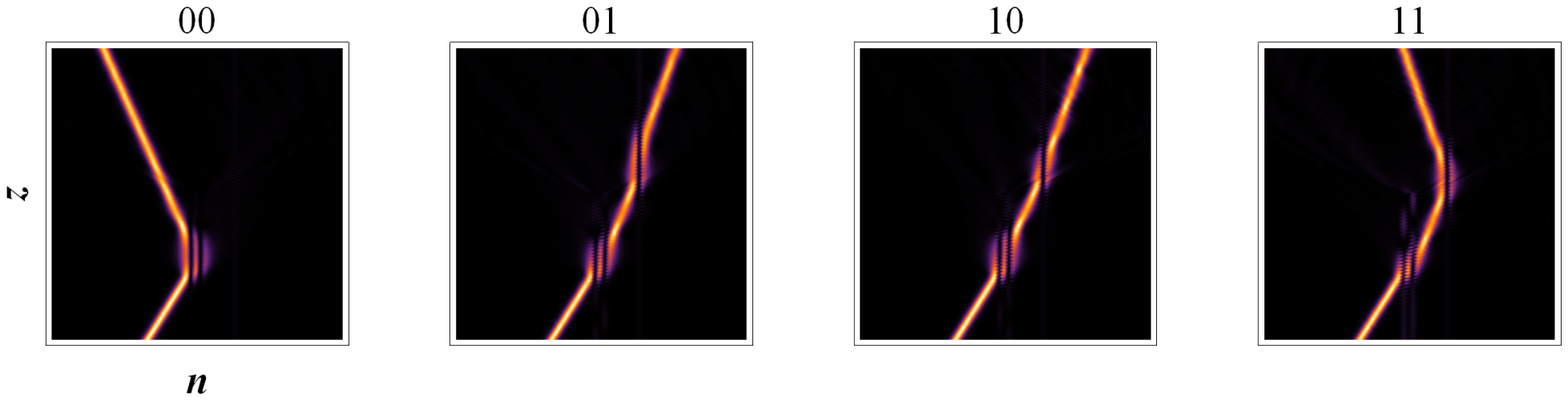}
        \caption{Spacio-temporal plots representing the function of XOR gate with initial soliton speed $v=0.22$. The two modified potential wells are separated by 4 waveguides, $\Delta n =4$, and take the same position as in the OR gate. The third potential well is placed at 15 waveguides from the centre. The control solitons are invisible for clear density plots. The parameter setting the power of the control solitons in the first two wells is $r=0.16$, as in the OR gate, while in the third well $p=0.1$. Time ranges from $z=0$ to $z=423$. The reflection for the (11) case is enhanced by cutting off the first two potential wells at $z=212$. Other parameters used are $c=1$ and $\mu=1.5$.}
        \label{fig:dp-XOR}
    \end{figure}

\begin{figure}[!h]
    \hspace*{-0.7cm} 
        \centering
        \includegraphics[width=0.8\linewidth]{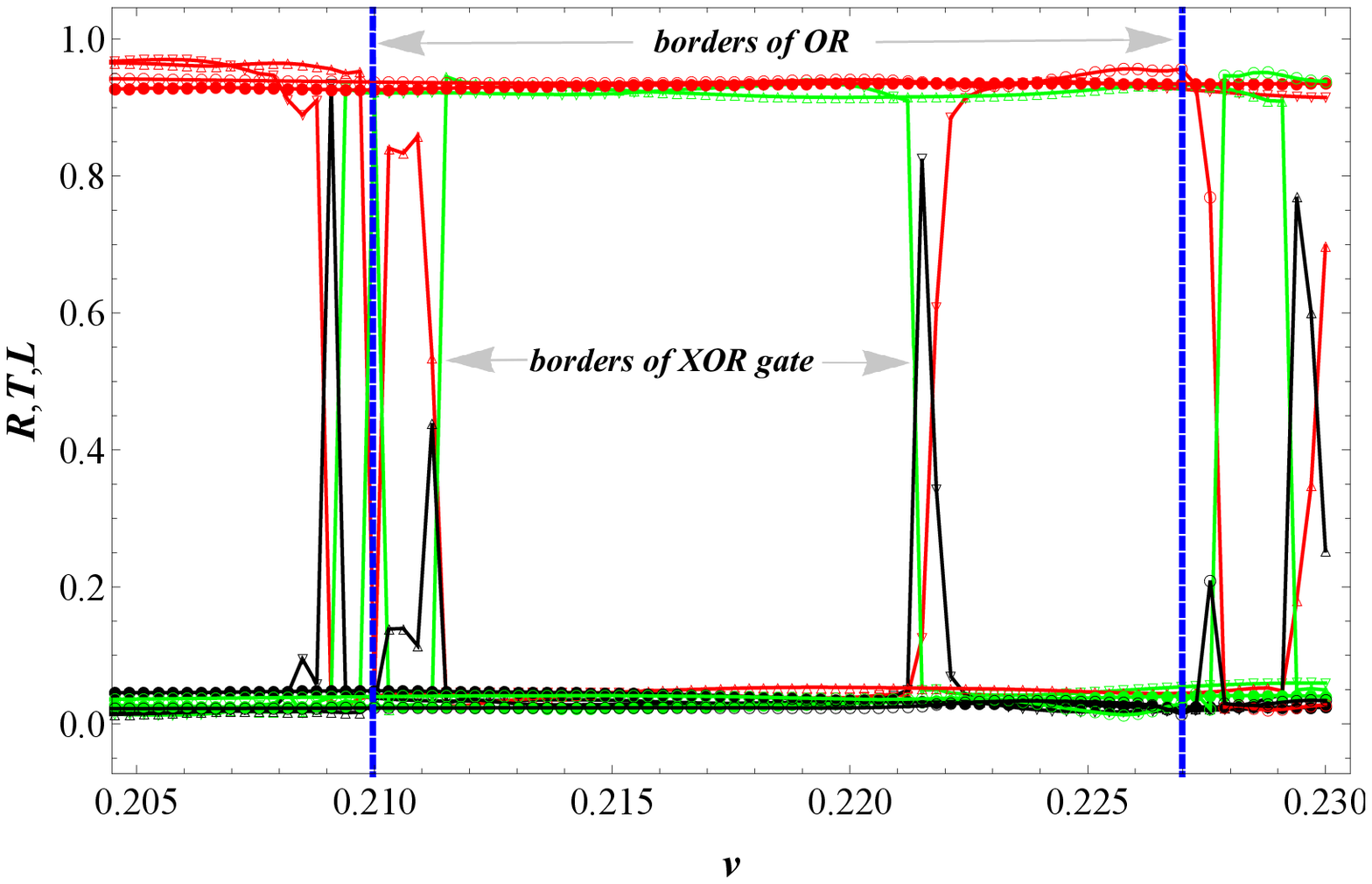}
        \caption{Transport Coefficients of XOR gate: Red curves correspond to reflection (R),
            green curves correspond to transmission (T), and black curves correspond to trapping (L). Filled circles correspond to the presence of control solitons in both wells (11). Up and lower triangles correspond to the presence of a control soliton in the left or right well (10 or 01), respectively. Empty circles correspond to the absence of control solitons from both wells (00). The working window of velocity for OR and XOR gate are considerably comparable. All parameters used here are the same as in Fig.\ref{fig:dp-XOR}.}
        \label{fig:TC-XOR}
    \end{figure}

\begin{figure}[!h]

\centering
\hspace{0.cm}\includegraphics[scale=0.53]{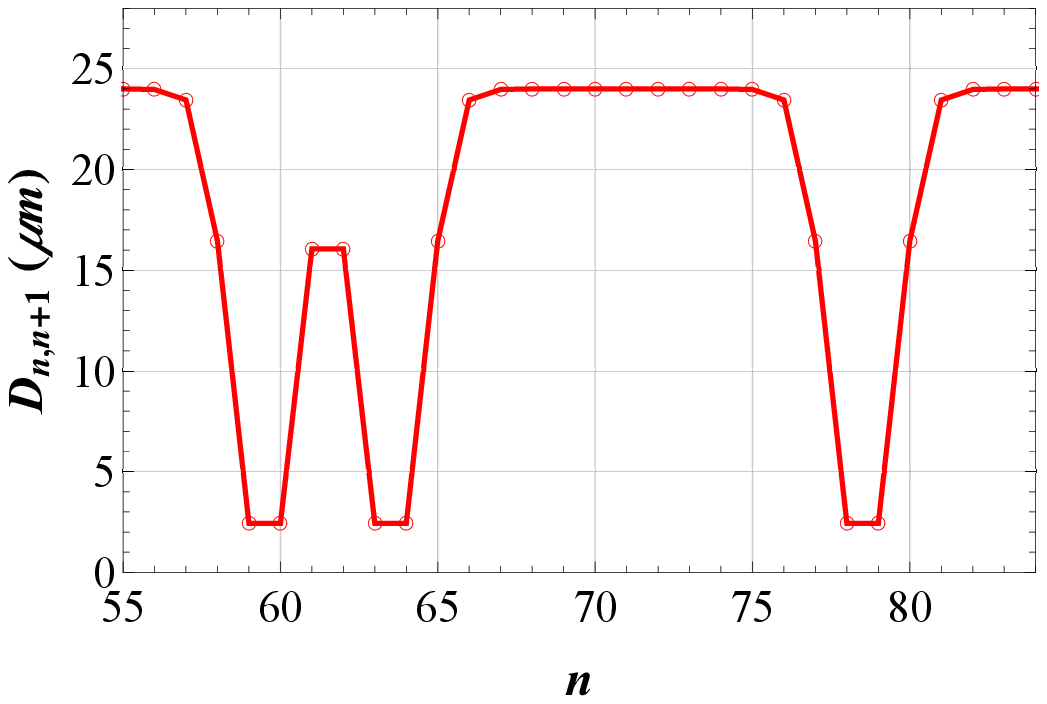}\,\,\,\,\,\,\,\,\,\,\,\,\,\,\,
\vspace{0.2cm}
\hspace{0.cm}\includegraphics[scale=0.55]{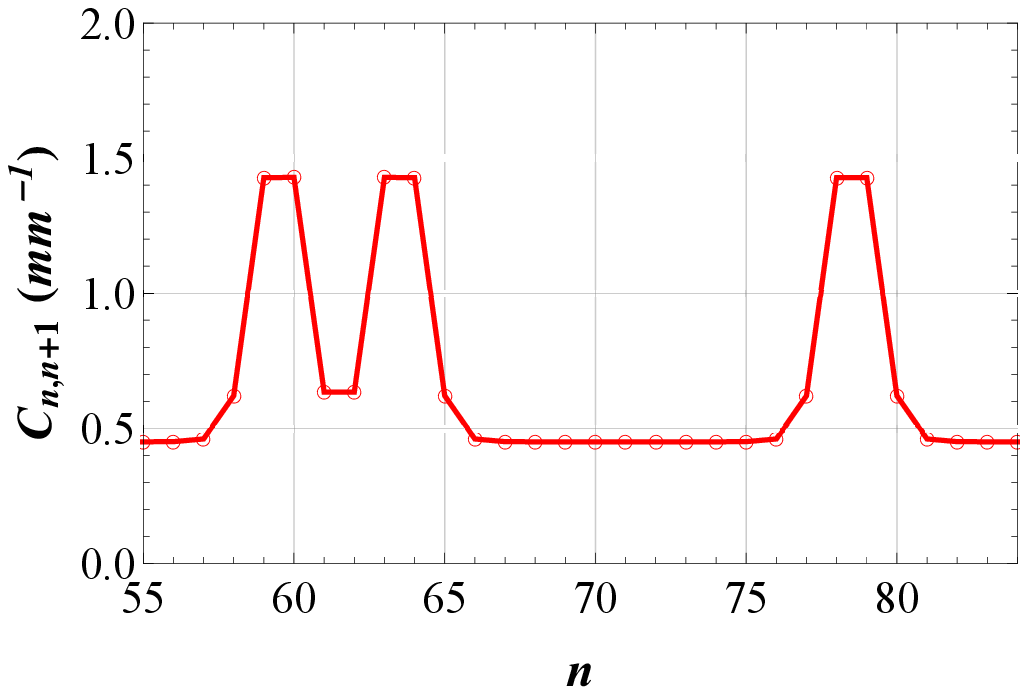} 

\centering
\includegraphics[scale=0.53]{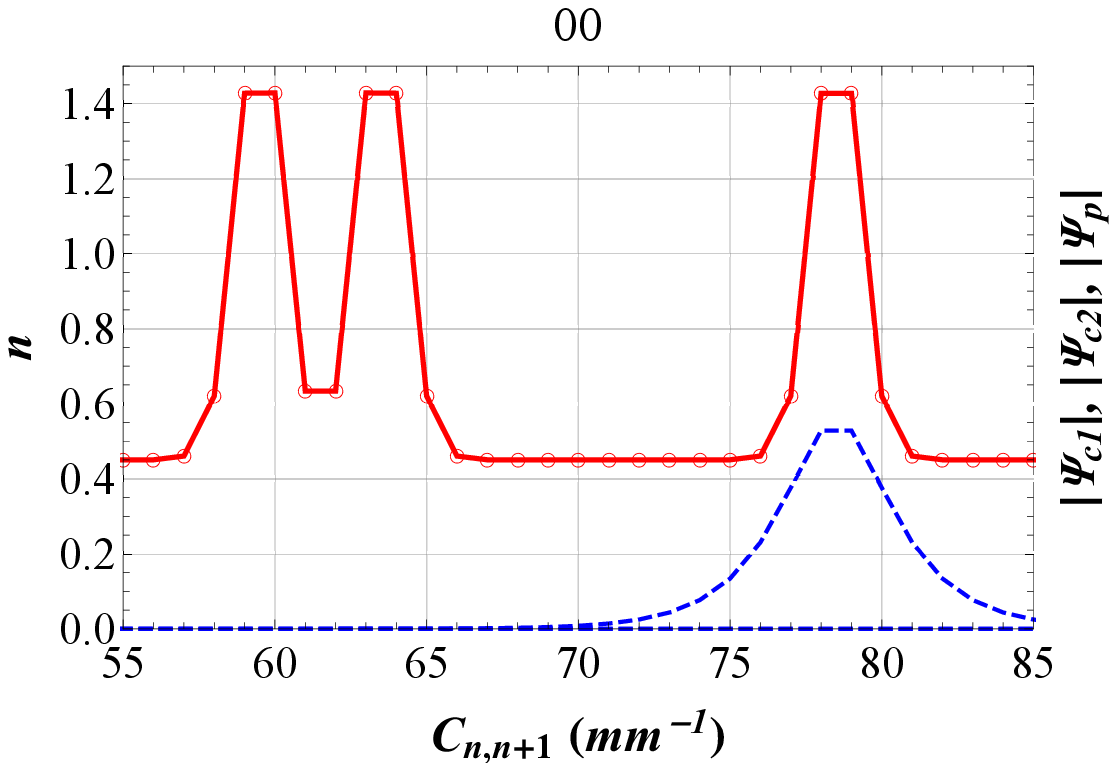}\hspace{0.5cm}
\vspace{0.2cm}
\includegraphics[scale=0.53]{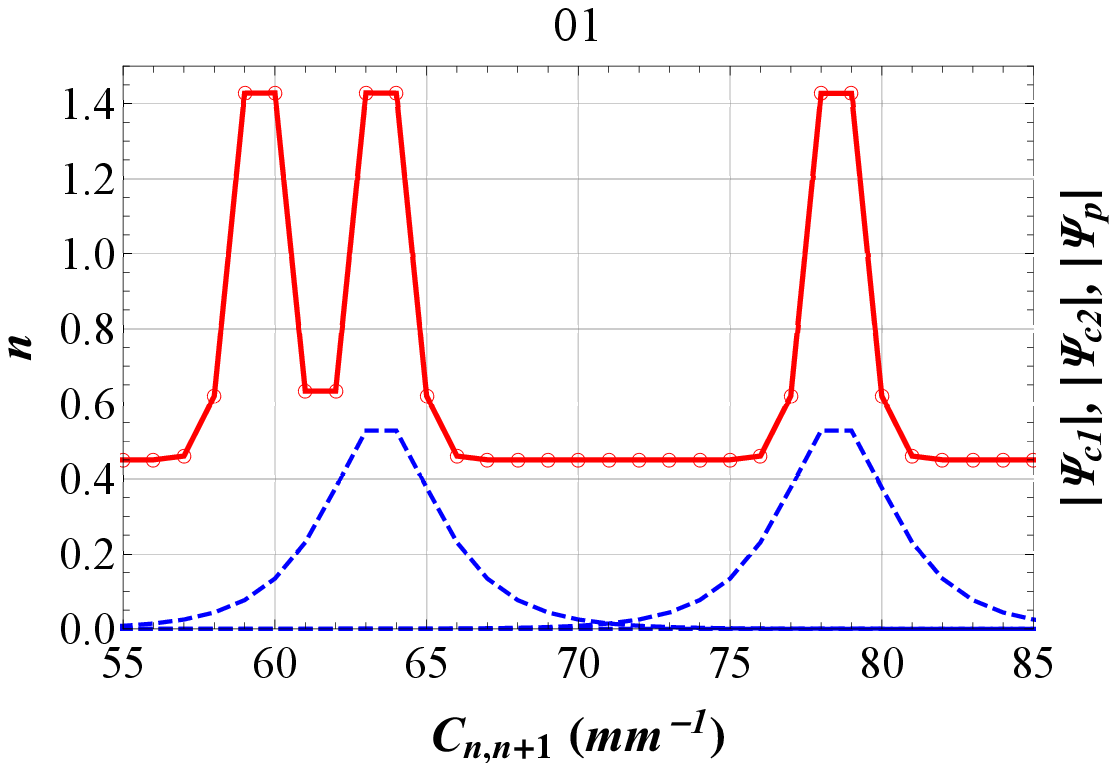}\hspace{0cm}

\centering
\includegraphics[scale=0.53]{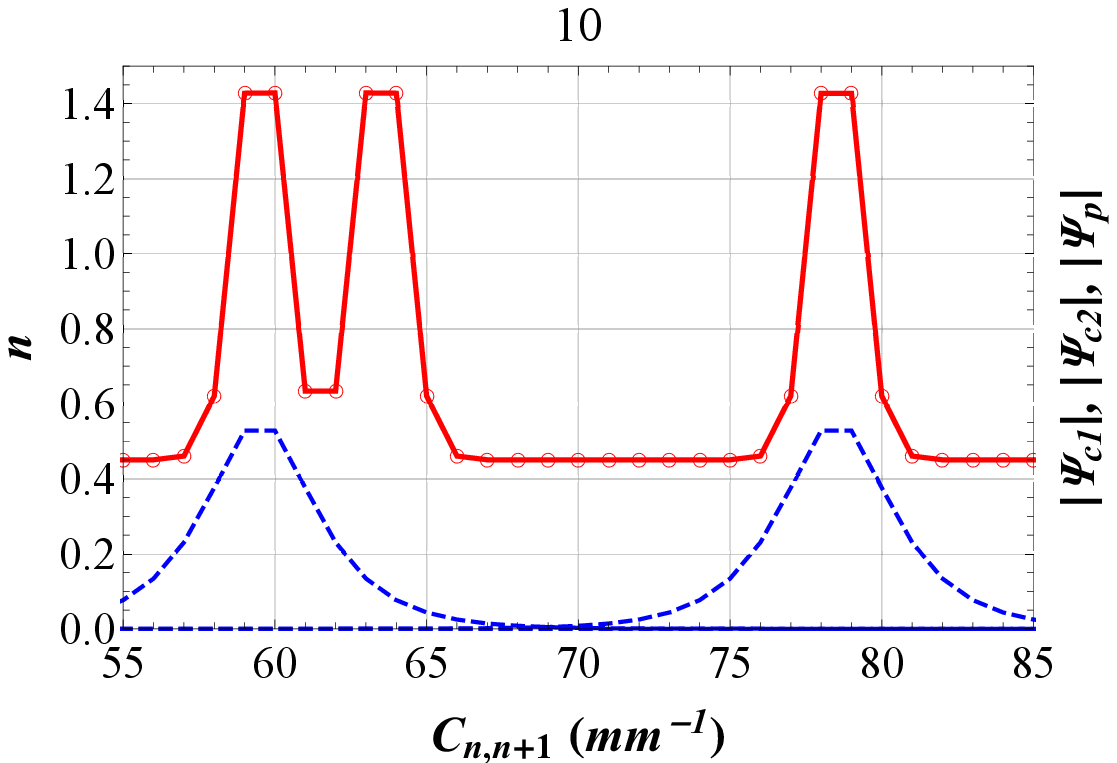}\hspace{0.5cm}
\includegraphics[scale=0.53]{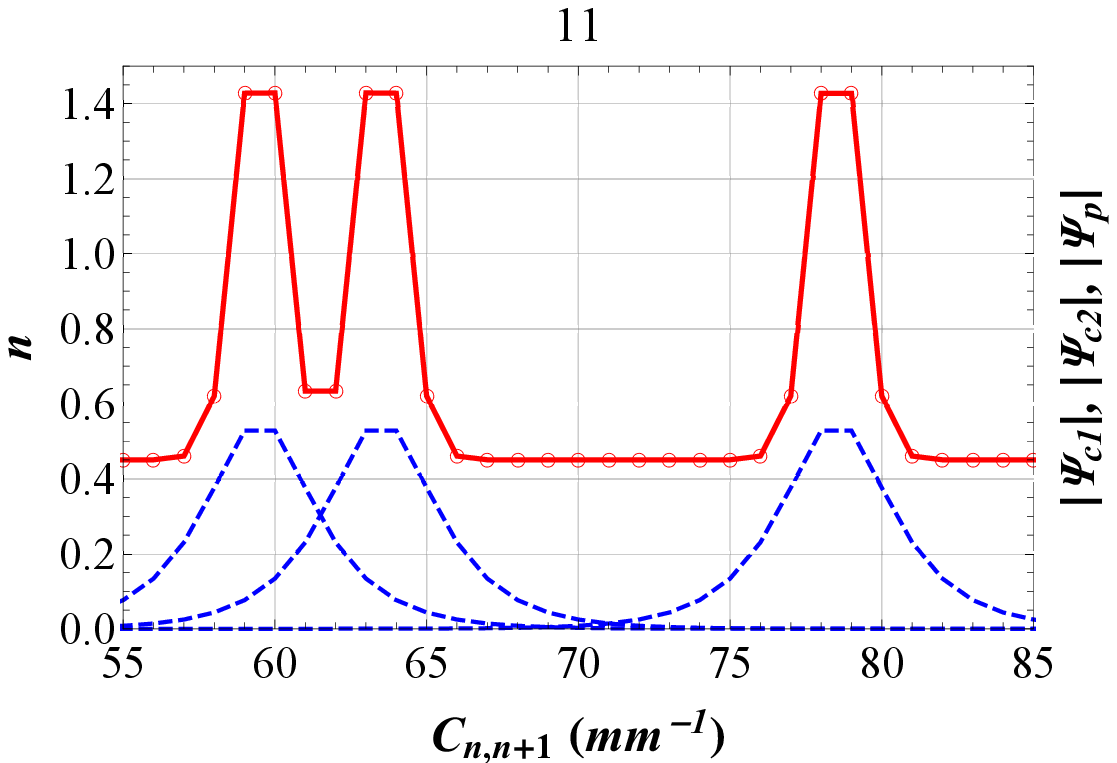}\hspace{0cm} 
\caption{Left subfigure in the upper panel shows waveguides separation, $D_{n,n+1}$.
The right subfigure in the upper panel shows waveguides coupling, $C_{n,n+1}$. The calibration values $D_0=24~\mu$m and $C_0=0.45 {~mm^{-1}}$ were taken from the experiment of \cite{2} for the $\lambda=543$ nm pulse. The lower panel consist of four subfigures corresponding to 00, 01, 10, and 11. The blue dashed line represents the intensity of the control solitons.}
        \label{fig:D&C}
    \end{figure}

    The modification we make on the OR gate is based on distinguishing
    the scattered solitons according to their center-of-mass speeds.
    The trajectories of discrete solitons are characterized by
    quick jumps between the waveguides (large speed) and temporary
    pinning at the waveguides (small speeds), as shown in
    Fig.~\ref{fig:benchmark}.
    The speed will then have two different values and an averaging procedure is therefore needed. We conduct this averaging by calculating the speed over a time interval corresponding to a number of
    waveguides. The result is shown in Fig.~\ref{fig:benchmark}.
    Applying the averaging procedure to all possible cases for the
    action of OR gate, we notice a clear difference in the speeds of the
    output soliton, as shown in Fig.~\ref{fig:speeds}. The free
    propagation of signal soliton, which we take as a benchmark, preserves the highest
    state of velocity among all. This is of course expected due to the absence of
    potential wells which are main sources of dissipation. The speed of the (11) case is noticeably lower than
    that of the (01) and (10) cases. The speed for the (00) case is
    negative since it is reflection. This difference in speeds is the
    key element for designing the XOR gate.
    We exploit this difference by introducing a third potential well and control soliton that
    reflect the (11)
    case but not the (01) and (10). This is possible due to the fact that the critical speed at which the solitons reflect or transmit can be controlled by the intensity of the control soliton. Tuning the intensity of the third control soliton, we managed to achieve the function of the XOR gate. Schematically, the XOR gate is represented by Fig.~\ref{fig:schematic-XOR}.\\
 \\The distribution of control solitons for the XOR gate are as follows: For the 00 gate, we have no control solitons in the input ports, for the 01 gate we have only one control soliton in the second input port, for the 10 gate, we have only one control soliton in the first input port, and for the 11 gate, we have two control solitons in the input ports. The third control soliton on the right side of the input ports with different control power is used particularly to reflect the (11) case in order to achieve the performance of XOR gate.\\
    In Fig.~\ref{fig:dp-XOR}, the desired performance is clearly seen to have been achieved. The (00) and (11) reflect while (01) and (10) transmit. The transport coefficients for the XOR gate
    are presented in Fig.~\ref{fig:TC-XOR}. It is seen from the figure
    that the working window of velocity for the XOR gate is reasonably
    comparable to and overlaps with the OR gate.
    For the feasibility of experimental realization, we plot the
    waveguides separation in $\mu m$ and the corresponding coupling
    strength in $mm^{-1}$ corresponding to the potential wells used in
    Fig.~\ref{fig:TC-XOR}, as calculated by using Eqs.~\eqref{Ceq}
    and \eqref{Deq} and shown in Fig.~\ref{fig:D&C}.
    
\section{One-bit half adder and full adder}
	\label{adderssec}
\begin{figure}[!h]
		\centering
			\includegraphics[scale=.06]{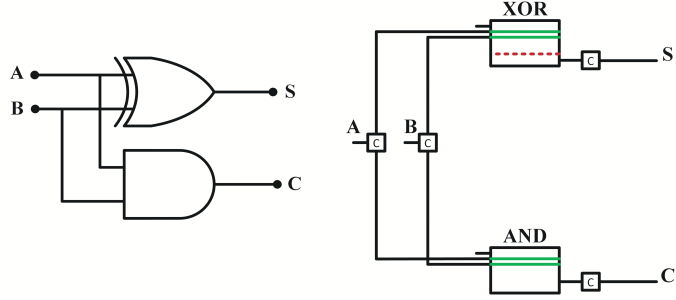}
			\vspace{0cm}
		\caption{Left subfigure shows the typical electronic scheme of half adder designed with the combination of AND and XOR gate. A and B are the two inputs while S (sum) and C (carry) are the outputs. Right subfigure shows our scheme of half adder designed with the discrete solitons in waveguide arrays. c in box is a converter modulates the intensity of a input soliton to match that of a control soliton.}
		\label{fig:halfadder}
	\end{figure}

\begin{figure}[!h]
	\hspace*{0.5cm} 
	\centering
		\includegraphics[scale=.08]{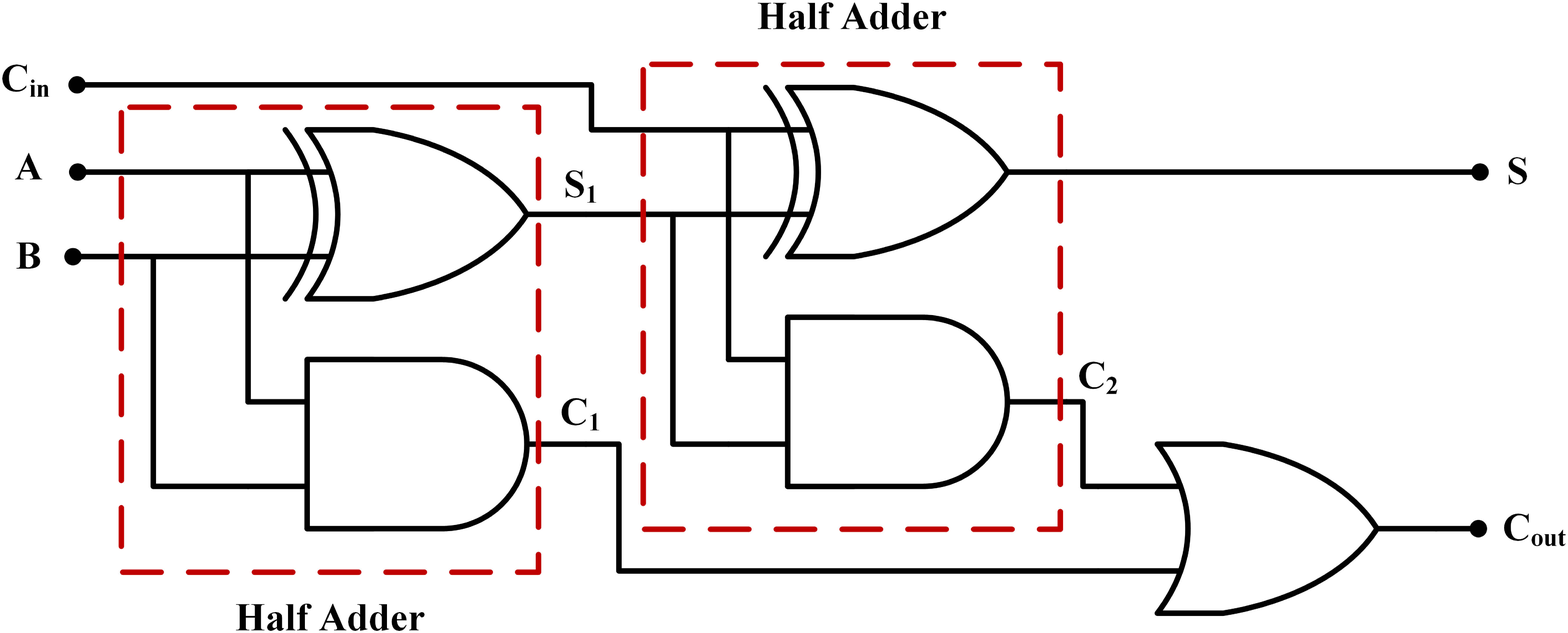}
		\vspace{0.7cm}
		\includegraphics[scale=.08]{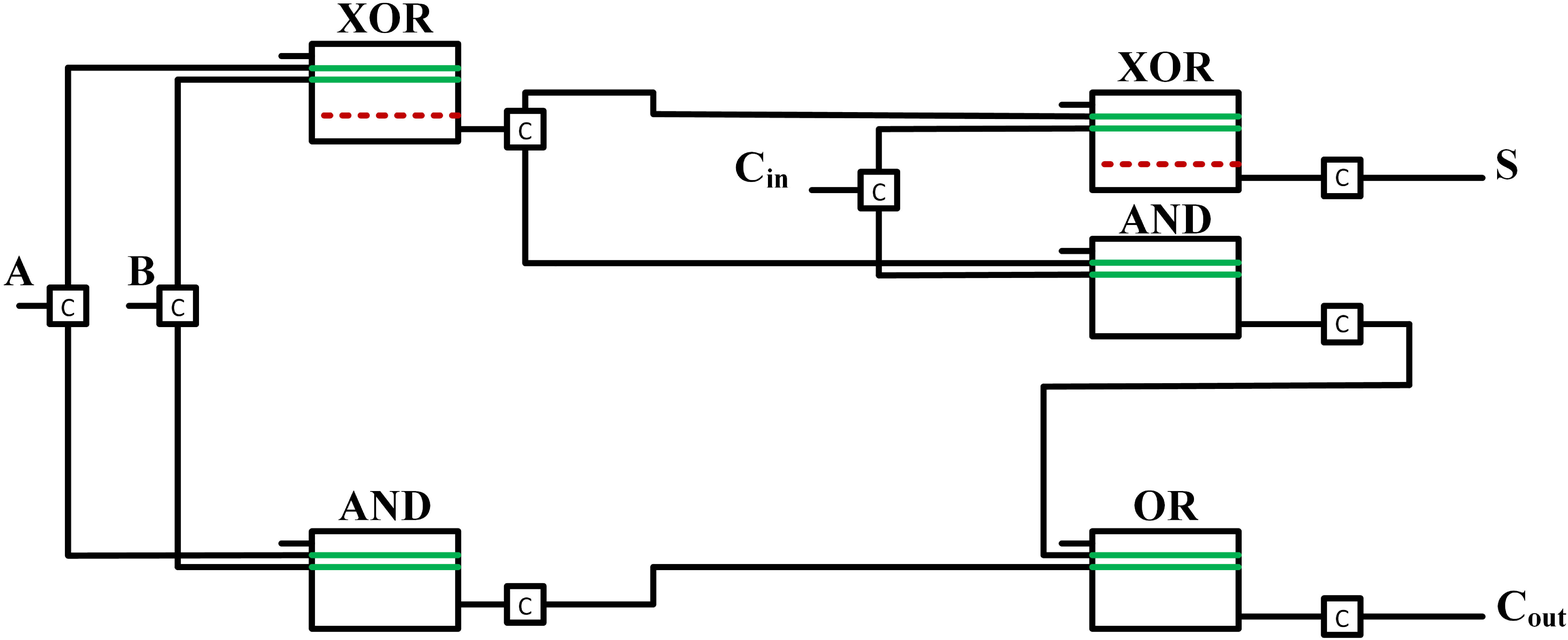}	
	\caption{Upper subfigure shows the electronic scheme of full adder designed with the combination of two half adders. A, B and C$_{in}$ are the three inputs while S (sum) and C$_{out}$ (carry) are the outputs. Lower subfigure represents our scheme of full adder designed with the combination of two half adders shown in Fig.\ref{fig:halfadder}.}
		\label{fig:fulladder}
	\end{figure}

	Adders are type of digital circuits designed to perform the addition
	of numbers. There are two types: half adders and full adders. As
	shown in the left subfigure in Fig.\ref{fig:halfadder}, the half
	adder has two inputs, assigned A and B, and two outputs S (Sum) and
	C (Carry). The half adder is able to add two single binary digits
	and provides two digit outputs (sum and carry). The C (carry) value
	is an AND output of the inputs A and B while the S (Sum) value is
	the output from XOR gate. Therefore, the common representation of
	the half adder simply uses an XOR logic gate and an AND logic gate \cite{haldadder}.\\
	\\Similarly, we design our scheme for the half adder with the use of
	discrete solitons in waveguides arrays, composed of an AND gate and
	an XOR gate. Our half adder takes two inputs, A and B, and delivers
	two outputs S (sum) and C (carry). A and B inputs are attached to
	the potential wells by injecting the control solitons. The output S
	is collected from the XOR gate upon the transmittance of the
	discrete soliton. In addition, the output C is taken from the AND gate upon
	the transmittance. 	In order to maintain the universal features of
	the signal soliton, a converter is introduced in our scheme. The purpose of the converter is to allow for devices connectivity and hence scalability. It refers to a pulse generator that produces an output signal that is always of the same characteristics regardless of the input signal. For instance, the input signal of a converter can be an output (reflected or transmitted) soliton. It takes the output signal of a device and converts it to an input signal to another device. Our half adder scheme is shown in Fig.\ref{fig:halfadder}.
	The full adder extends the concept of the half adder by providing an
	additional carry-in (C$_{in}$) input as described in Ref.~\cite{fulladder}. It adds three single-digit
	binary numbers, two inputs and a carry-in bit. The third input makes
	it eligible for scaling $n$-bits. The full adder outputs two
	numbers, a Sum and a Carry bit. Upper subfigure in
	Fig.\ref{fig:fulladder} shows the electronic scheme of full adder
	with three inputs: A, B and C$_{in}$, which adds the three input
	numbers and generates a Carry out ($C_{out}$) and a Sum ($S$). As a
	result, placing two half adders together with the use of an OR gate
	results in a one-bit full adder. In order to add more $n$-bits,
	$n$-full adders are connected in cascade to perform the addition.
	We construct the full adder by following the electronic scheme in structure. It has three inputs, A, B, and C$_{in}$, and two outputs the S (sum) and C (carry) as shown in the lower subfigure in Fig.\ref{fig:fulladder}. A converter is also introduced in this scheme. 
\section{Conclusion} \label{concsec}
We have shown that by modifying the setup of the previously-proposed OR gate in a waveguide array, an XOR gate can be obtained. The modification includes the introduction of a third potential well in addition to the two potential wells in the OR gate. A control soliton is injected into the new potential well. The role of the new potential well and its control soliton is to disperse the (11) output signal from the (10) and (01) outputs exploiting a difference in their center-of-mass speed. We found this to be possible within a finite width of a velocity of incidence that is comparable with that of the OR gate. To facilitate the experimental realization, the profiles of the waveguides separations have been calculated in real units, as given by Fig. \ref{fig:D&C}. We have also shown how the AND, OR, and XOR gates can be connected to result in half and full adders. We believe this proposal will be a useful step towards achieving all-optical data processing.
\section*{Acknowledgment} The authors acknowledge the support of UAE University through grants UAEU-UPAR(4) 2016 and UAEU-UPAR(6) 2017.


\begin{thebibliography}{1}
\newcommand{\enquote}[1]{``#1''}
\bibitem{books1} A. Hasegawa and Y. Kodama, ``Solitons in optical communications", Oxford: Oxford Univ. Press, (1995).

\bibitem{books2} L. Mollenauer and J. Gordon, ``Solitons in optical fibers", Boston: Acadamic Press, (2006).

\bibitem{books3} N. Akhmediev and A. Ankiewicz, ``Solitons: Nonlinear Pulses and Beams", London: Chapman and Hall, (1997).

\bibitem{books4} G. Agrawal, ``Nonlinear fiber optics", 3rd ed, San Diego: Academic Press, (2001).

\bibitem{gates1}R. Keil, M. Heinrich, F. Dreisow, T. Pertsch, A. Tünnermann, S. Nolte, D. Christodoulides and A. Szameit, ``All-optical routing and switching for three-dimensional photonic circuitry", Sci. Rep., Vol. \textbf{1}, 94, (2011).

\bibitem{gates2} A. Politi, M. Cryan, J. Rarity, S. Yu and J. O'Brien, ``Silica-on-Silicon Waveguide Quantum Circuits", Science, Vol. \textbf{320}, 5876, pp. 646 - 649 (2008).

\bibitem{gates3} J. Sabini, N. Finlayson and G. Stegeman, ``All optical switching in non linear X junctions", Appl. Phys. Lett., Vol. \textbf{55}, 12, pp. 1176 - 1178 (1989).

\bibitem{gates4} P. Chu, Y. Kivshar, B. Malomed, G. Peng and M. Quiroga-Teixeiro, ``Soliton controlling, switching, and splitting in nonlinear fused-fiber couplers", J. Opt. Soc. Am. B, Vol., \textbf{12}, 5, pp. 898 - 903 (1995).
 
\bibitem{gates5}  M. Peccianti, C. Conti, G. Assanto, A. De Luca, and C. Umeton, ``All-optical switching and logic gating with spatial solitons in liquid crystals", Appl. Phys. Lett., Vol. \textbf{81}, 18, pp. 3335 - 3337 (2002).

\bibitem{gates6}  A. Piccardi, A. Alberucci, U. Bortolozzo, S. Residori, and G. Assanto, ``Soliton gating and switching in liquid crystal light valve", Appl. Phys. Lett., Vol. \textbf{96}, 7, pp. 071104 - 071104 (2010).

\bibitem{gates7} W. Krolikowski and Y. Kivshar, ``Soliton-based optical switching in waveguide arrays", J. Opt. Soc. Am. B, Vol. \textbf{13}, 5, pp. 876 - 887 (1996).

\bibitem{gates8}  Y. Wu, ``All-optical logic gates by using multibranch waveguide structure with localized optical nonlinearity", IEEE J. Sel. Top. Quantum Electron, Vol. \textbf{11}, 2, pp. 307 - 312 (2005).

\bibitem{gates9}  Y. Wu, ``New all-optical switch based on the spatial soliton repulsion", Opt. Express, Vol. \textbf{14}, 9, pp. 4005 - 4012 (2006).

\bibitem{gates10}  Y. Wu, M. Huang, M. Chen, R. Tasy, ``All-optical switch based on the local nonlinear Mach-Zehnder interferometer", Opt. Express, Vol. \textbf{15}, 16, pp. 9883 - 9892 (2007).

\bibitem{gates11}  K. M. Aghdami, M. Golshani, and R. Kheradmand, ``Two-dimensional discrete cavity solitons: switching and all-optical gates", IEEE Photon. J., Vol. \textbf{4}, 4, pp. 1147 - 1154 (2012).

\bibitem{gates12} D. Christodoulides and E. Eugenieva, ``Blocking and Routing Discrete Solitons in Two-Dimensional Networks of Nonlinear Waveguide Arrays", Phys. Rev. Lett., Vol. \textbf{87}, 23, (2001).

\bibitem{gates13}  J. Scheuerand, M. Orenstein, ``All-optical gates facilitated by soliton interactions in a multilayered Kerr medium", J. Opt. Soc. Am. B, Vol. \textbf{22}, 6, pp. 1260 - 1267 (2005).

\bibitem{books5} J. Taylor, ``Optical Solitons -Theory and Experiment", Cambridge: Cambridge Univ. Press, (1992).

\bibitem{fedor1} M. Stratmann, T. Pagel, and F. Mitschke, ``Experimental observation of temporal soliton molecules", Phys. Rev. Lett., Vol. \textbf{95}, 14, pp. 143902-1 - 143902-4 (2005).

\bibitem{fedor2} A. Hause, H. Hartwig, M. Bohm, and F. Mitschke, ``Binding mechanism of temporal soliton molecules", Phys. Rev. A, Vol. \textbf{78}, 6, pp. 063817-1 - 063817-9 (2008).

\bibitem{fedor3} A. Hause, H. Hartwig, B. Seifert, H. Stolz, M. Bohm, and F. Mitschke, ``Phase structure of soliton molecules", Phys. Rev. A,  Vol. \textbf{75}, pp. 063836-1 - 063836-2 (2007).

\bibitem{usamanjp} U. Al Khawaja and H. Stoof, ``Formation of matter-wave soliton molecules", New J. Phys., Vol. \textbf{13}, 8, pp. 085003-1 - 085003-11 (2011).

\bibitem{andrey} A. Sukhorukov, ``Reflectionless potentials and cavities in waveguide arrays and coupled-resonator structures", Opt. Lett., Vol. \textbf{35}, 7, pp. 989 - 991 (2010).

\bibitem{lepri} S. Lepri and G. Casati, ``Asymmetric Wave Propagation in Nonlinear Systems", Phys. Rev. Lett., Vol. \textbf{106}, 16, pp. 164101-1 - 164101-4 (2011).

\bibitem{usamaandrey} U. Al Khawaja and A. Sukhorukov, ``Unidirectional flow of discrete solitons in waveguide arrays", Opt. Lett., Vol. \textbf{40}, 12, pp. 2719 - 2722 (2015).

\bibitem{submitted} U. Al Khawaja, S. Al-Marzoug, and H. Bahlouli, ``Unidirectional flow of discrete solitons in optical waveguide arrays with modulated nonlinearity", \textbf{submitted} Jan. 2020.
     
\bibitem{usamaasad} U. Al Khawaja and M. Asad-uz-zaman, ``Directional flow of solitons with asymmetric potential wells: Soliton diode", EPL (Europhysics Letters), Vol. \textbf{101}, 5, pp. 50008-1 - 50008-7 (2013).

\bibitem{usamayuri} U. Al Khawaja, S. Al-Marzoug, H. Bahlouli and Y. Kivshar, ``Unidirectional soliton flows in PT-symmetric potentials", Phys. Rev. A, Vol. \textbf{88}, 2, pp. 023830-1 - 023830-6 (2013).

\bibitem{recentpre} M. Alotaibi, S. Al-Marzoug, H. Bahlouli and U. Khawaja, ``Unidirectional flow of solitons with nonlinearity management", Phys. Rev. E, Vol. \textbf{100}, 4, pp. 042213 (2019).

\bibitem{1} U. Al Khawaja, S. Al-Marzoug and H. Bahlouli, ``All-optical switches, unidirectional flow, and logic gates with discrete solitons in waveguide arrays", Opt. Express, Vol. \textbf{24}, 10, pp. 11062 - 11074 (2016).

\bibitem{rev} I. Garanovich, S. Longhi, A. Sukhorukov and Y. Kivshar, ``Light propagation and localization in modulated photonic lattices and waveguides", Physics Reports, Vol. {\bf518}, 1-2, pp. 1 - 79 (2012).

\bibitem{AL} M. Ablowitz, Y. Ohta and A. David Trubatch, ``On discretizations of the vector nonlinear Schr\"odinger equation", Phys. Lett. A, Vol.\textbf{ 253}, 5-6, pp. 287 - 304 (1999).

\bibitem{2} A. Szameit, F. Dreisow, T. Pertsch, S. Nolte and A. T\"unnermann, ``Control of directional evanescent coupling in fs laser written waveguides", Opt. Express, Vol. \textbf{15}, 4, pp. 1579 - 1587 (2007).

\bibitem{expt2} M. Bellec, G. Nikolopoulos and S. Tzortzakis, ``Faithful communication Hamiltonian in photonic lattices", Opt. Lett., Vol. \textbf{37}, 21, pp. 4504 - 4506 (2012).

\bibitem{Boris} J. Cuevas, G. James, P.G. Kevrekidis, B.A. Malomed and B. Sanchez-Rey, ``Approximation of solitons in the discrete NLS equation", J. Nonlinear Math. Phys, Vol. \textbf{15}, 3, pp. 124 - 136 (2008).

\bibitem{haldadder} E. Yaghoubi, L. A. Bakhtiar, A. Adami, S. Hamidi and M. Hosseinzadeh, ``All optical OR/AND/XOR gates based on nonlinear directional coupler", J. Opt., Vol \textbf{43}, 2, pp. 146 - 153, (2014). 

\bibitem{fulladder} E. R. Pashaki and M. Shalchian, ``Design and simulation of an ultra-low power high performance CMOS logic: DMTGDI", Integration, Vol.\textbf{55}, pp. 194 - 201, (2016). 

\end{thebibliography}
\end{document}